\documentclass[aps,prb,twocolumn,superscriptaddress,longbibliography, citeautoscript]{revtex4-1}%
\setcitestyle{numbers,square}
\usepackage{graphicx}
\usepackage{float}
\usepackage{dcolumn}
\usepackage{bm,color}
\usepackage{amsmath,amssymb,dsfont,amstext,amsfonts}
\usepackage{extarrows}
\usepackage[colorlinks=true,linkcolor=blue,urlcolor=blue,citecolor=blue]{hyperref}
\usepackage{xcolor}
\usepackage{wasysym}
\usepackage{mathtools}
\usepackage{bbold} 
\usepackage{microtype}
\usepackage{orcidlink}
\usepackage[english]{babel}
\usepackage{braket} 

\usepackage[normalem]{ulem} 
\usepackage{color}

\let\oldciteauthor=\citeauthor
\def\citeauthor#1{\hypersetup{citecolor=blue}\oldciteauthor{#1}}

\let\oldciten=\onlinecite
\def\onlinecite#1{\hypersetup{citecolor=blue}\oldciten{#1}}

\let\oldcite=\cite
\def\cite#1{\hypersetup{citecolor=blue}\oldcite{#1}}

\newcommand{\pos}{\boldsymbol{r}}

\begin{document}
\title{Aharonov-Bohm oscillations and perfectly transmitted mode in amorphous topological insulator nanowires}

\author{Miguel F.\ Martínez\orcidlink{0000-0002-4281-9861}}
\affiliation{Department of Physics, KTH Royal Institute of Technology, 106 91, Stockholm, Sweden}

\author{Adolfo G.\ Grushin\orcidlink{0000-0001-7678-7100}}
\affiliation{Univ.~Grenoble Alpes, CNRS, Grenoble INP, Institut Néel, 38000 Grenoble, France}
\affiliation{Donostia International Physics Center (DIPC), Paseo Manuel de Lardizábal 4, 20018 Donostia-San Sebastián, Spain}
\affiliation{IKERBASQUE, Basque Foundation for Science, Maria Diaz de Haro 3, 48013 Bilbao, Spain}

\author{Jens H.\ Bardarson\orcidlink{0000-0003-3328-8525}}
\affiliation{Department of Physics, KTH Royal Institute of Technology, 106 91, Stockholm, Sweden}

\begin{abstract}
Crystalline topological insulator nanowires with a magnetic flux threaded through their cross section display Aharanov-Bohm conductance oscillations.
A characteristic of these oscillations is the perfectly transmitted mode present at certain values of the magnetic flux, due to the appearance of an effective time-reversal symmetry combined with the topological origin of the nanowire surface states.
In contrast, amorphous nanowires display a varying cross section along the wire axis that breaks the effective time-reversal symmetry.
In this work, we use transport calculations to study the stability of the Aharanov-Bohm oscillations and the perfectly transmitted mode in amorphous topological nanowires.
We observe that at low energies and up to moderate amorphicity the transport is dominated, as in the crystalline case, by the presence of a perfectly transmitted mode. 
In an amorphous nanowire the perfectly transmitted mode is protected by chiral symmetry or, in its absence, by a statistical time-reversal symmetry.
At high amorphicities the Aharanov-Bohm oscillations disappear and the conductance is dominated by nonquantized resonant peaks.
We identify these resonances as bound states and relate their appearance to a topological phase transition that brings the nanowires into a trivial insulating phase. 
\end{abstract}

\maketitle
\section{Introduction}
A strong $3d$ topological insulator in the quasi-$1d$ geometry of a nanowire behaves differently than in a $3d$ geometry.
In $3d$ the energy spectrum hosts an odd number of Dirac cones at the surface~\cite{zhang2009} and is thus gapless.
The electronic states at the surface are spin-momentum locked and protected from localization by the interplay of time-reversal symmetry and topology~\cite{Bardarson.2007,Nomura.2007,Kane2010, moore2010, XiaoLiang2011,Jens2013}.
In a topological nanowire the energy spectrum is gapped, and there is an even number of modes~\footnote{Throughout the manuscript we use the term mode to refer to electronic states with a particular sign of momentum. For example, if there exists one electronic state with energy $E$ and $k<0$ and another state with the same energy and $k>0$, there is only one mode as for each sign of momentum there is only one state.} at every energy~\cite{Franz2010, Vishwanath2009, Mirlin2010, Jens2010, Jens2013,Ilan2018}.
The gap is due to the spin of the electrons acquiring a $\pi$ Berry phase when taking a closed trajectory around the circumference of the nanowire~\cite{Mirlin2010, Franz2010, Jens2010, Vishwanath2010}.
Threading a magnetic flux $\phi$ through the cross section of the nanowire allows to change the number of modes from even to odd.
At $\phi=\phi_0/2$ and up to integer multiples of the flux quantum $\phi_0=e/h$, the gap closes, restoring a Dirac cone and resulting in an odd number of modes at any Fermi energy.
This occurs because the Aharanov-Bohm phase exactly cancels the spin Berry phase at these particular values of the magnetic flux~\cite{Jens2010}.

The interplay between the number of modes and the time-reversal symmetry has consequences for the conductance of the nanowire~\cite{Ilan2018}.
Although the time-reversal symmetry is broken due to the magnetic flux, it is effectively recovered at the surface for $\phi=(2n+1) \phi_0/2$.
Intuitively, the symmetry results from electrons traveling in opposite directions around the circumference of the nanowire, picking up the same phase factor.
The odd number of modes, together with time-reversal symmetry, implies that at $\phi=(2n+1)\phi_0 /2$  at least one reflection eigenvalue vanishes due to the antisymmetry of the reflection matrix~\cite{Jens2008}.
A perfectly transmitted mode protected from backscattering appears then in the nanowire.
The conductance displays Aharanov-Bohm oscillations with varying magnetic flux and a quantized value at $\phi=(2n+1)\phi_0/2$~\cite{Jens2010, Jens2013, Vishwanath2010}.
The oscillations are a signature of the topological nature of the surface state in a $3d$ topological insulator~\cite{Ilan2018}.

Topological nanowires have been studied in experiments~\cite{peng2010, xiu2011, Dufouleur2013, hong2014, cho2015, Jauregui2016, Dufouleur2017, Ziegler2018, Munning2021,  Feng2025} and are potential candidates for technological applications, for example as electrical interconnects~\cite{gilbert2021, gupta2014, philip2016, rocchino2024, Khan2025,Gilbert2025,Cheon2025,Soulie2024}.
The resistivity of traditional copper-based interconnects increases due to finite-size effects as their dimensions are reduced to the nanoscale~\cite{Engelhart2002, steinhogl2005}.
The absence of backscattering of topological surface states in nanowires may circumvent this problem.
This strategy is already being investigated in topological-insulator~\cite{gupta2014, philip2016} and Weyl-semimetal~\cite{rocchino2024, Khan2025,Cheon2025, Kaladzhyan2019} nanowires.
Another possible application is in the context of quantum computation.
Analogously to semiconducting nanowires with strong spin-orbit coupling~\cite{Oreg2010, Sarma2010, alicea2011},
topological nanowires in contact with a superconducting substrate can be brought to a topological superconducting state with a Majorana zero mode at their ends~\cite{kitaev2001, Cook2011, Cook2012, Juan2014, Manousakis2017, Jens2019, bai2022,Feng2025,Kaufhold2025}, which can be used as a topological qubit.

The feasibility of these technological applications depends on the properties of topological nanowires being robust to, among other effects, disorder and structural imperfections that are common in real materials. 
Conversely, it is important to find materials that provide versatile and scalable platforms to create these nanowires.
Amorphous topological materials~\cite{Agarwala2017, xiao_photonic_2017,mitchell_amorphous_2018, poyhonen_amorphous_2018,mansha_robust_2017, costa_toward_2019, Agarwala2020, Marsal2020, Corbae2021,spillage_2022, Adolfo2022, Julia2023} are one such platform, as they are easier to grow than perfectly crystalline structures.
Recent experimental progress in which the surface state of amorphous $\mathrm{Bi}_2\mathrm{Se}_3$~\cite{TIamorphousThinFilm,Banerjee:2017jd} has been observed by ARPES measurements~\cite{corbae2023, Adolfo2024} exemplifies their effectiveness as candidates for topological materials.
When other topological materials are grown amorphous they still display transport effects often attributed to a non-zero Berry curvature~\cite{DC2018,Ramaswamy:2019bp,Bouma2020,Li2021aWTex,Molinari2023,Fujiwara2023kagome,KarelAmorphousBerryCMG}.
Another example is the recent observation that electrical resistivity in thin films of noncrystalline CoSi and NbP is lower than in conventional metals~\cite{rocchino2024,Khan2025}, potentially useful for nanoelectronic applications.

Although promising, nanowires with an underlying amorphous structure can break the effective time-reversal symmetry of the surface due to the variation of their cross section~\cite{Jens2020}.
It is thus pertinent to ask under which conditions the Aharanov-Bohm oscillations and perfectly transmitted mode survive in an amorphous nanowire.
This question has been discussed for disordered nanowires~\cite{Jens2010, Jens2013, Jens2017}, but
studies on amorphous nanowires remain scarce.
Ref.~\onlinecite{Jens2020} studied nanowires with longitudinal ripples using a surface theory approach, Ref.~\onlinecite{Kozlovsky2020} studied cone-shaped wires, while Ref.~\onlinecite{mal2024} focused on transport calculations in a tight-binding model of a diamond lattice with connectivity defects.

In this work, we perform transport calculations on amorphous nanowires, both in the case of a constant cross section along the axis and in a fully amorphous case where there is no notion of a well-defined cross section. 
We use a tight-binding model with a crystalline limit describing the family of $\mathrm{Bi}_2\mathrm{Se}_3$ materials,
one of the most prominent examples of 3$d$ topological insulators both in crystalline and amorphous forms.
We investigate the presence of a perfectly transmitted mode and Aharanov-Bohm oscillations, and relate their robustness to the topological phase of the bulk of the nanowires by using local topological markers~\cite{Kitaev20062,Resta2011,LORING2015, Liu2018, loring2019, Julia2022, Julia2024}.
%

\section{Model}
We consider two types of amorphicity: layered amorphicity and full amorphicity.
We define a layered-amorphous nanowire as a stack of identical two-dimensional amorphous layers stacked along the $z$-direction.
The $(x, y)$-site coordinates are drawn from a normal distribution of standard deviation $w$ and mean value at the position of a square lattice.
In fully amorphous nanowires, all three site coordinates are drawn from a normal distribution with standard deviation $w$ and mean value at the site of a cubic reference lattice.
The reference lattice's cross section is spanned by $N_x$ and $N_y$ sites, and has $L$ sites along the $z$-direction.
We refer to $w$ as the \textit{amorphicity} of the lattice.

We endow the amorphous nanowires with a Hamiltonian adapting Fu and Berg's model~\cite{zhang2009, Zhang2010, Berg2010} for a $3d$ topological insulator with spin-$1/2$ and two $p_z$ orbitals to the noncrystalline case.
The resulting Hamiltonian reads
\begin{equation}
\label{eq: Amorphous hamiltonian}
\begin{split}
    &\quad \qquad \hat{\mathcal{H}} =\sum_{\pos \pos'} \hat{c}^\dagger_{\pos} h_{
    \pos \pos'}  \hat{c}_{\pos '},\\
    h_{
    \pos \pos'}& = (\epsilon \tau_1 - E_F\tau_0)\delta_{\pos\pos'} - t_{
    \pos \pos'}\tau_1  -\dfrac{i}{2}\eta_{
    \pos \pos'} \tau_2 \cos \theta\\
    &+\dfrac{i}{2} \lambda_{
    \pos \pos'}  \tau_3(\sigma_2\cos \varphi \sin \theta - \sigma_1 \sin \varphi \sin\theta),
\end{split}
\end{equation}
where $\epsilon$ is an onsite energy, $t_{\pos\pos'}$ and $\eta_{\pos\pos'}$ are hopping amplitudes, $\lambda_{\pos\pos'}$ controls the strength of the spin-orbit coupling, $\delta_{\pos\pos'}$ is the Dirac delta function, and $E_F$ is the Fermi energy.
The Pauli matrices $\tau_i$ and $\sigma_i$ act on the orbital and spin degrees of freedom, respectively.  
The fermionic operators at position $\pos$ are defined as $\hat{c}_{\pos} = (\hat{a}_{\pos\uparrow}, \hat{a}_{\pos\downarrow}, \hat{b}_{\pos\uparrow}, \hat{b}_{\pos\downarrow})^T$, where $\hat{a}, \hat{b}$ refer to each of the two orbitals.
The angles $\varphi$ and $\theta$ are the spherical coordinate angles between $\pos$ and $\pos'$ \footnote{Note that there is an implicit assumption of a global reference frame, as we fix the $x$-, $y$- and $z$-coordinates to be the same for each lattice site. This assumption can be relaxed by introducing a rotation matrix choosing an arbitrary reference frame for each site.}.
The hopping amplitudes and spin-orbit coupling depend on the position of the sites as
\begin{equation}
\label{eq: amorphous generalisation}
 f_{\pos\pos'} =f\Theta(\vert \pos-\pos' \vert - r_0)e^{-\vert \pos-\pos' \vert + 1} \quad \text{for} \quad \pos\neq\pos'
\end{equation}
where $f\in \lbrace t, \eta, \lambda\rbrace$, and vanish for $\pos=\pos'$.
The cutoff length $r_0$ determines the maximum distance between sites connected by the Hamiltonian, and we have set the lattice constant of the reference cubic lattice $a=1$.

The model has a time-reversal symmetry
\begin{equation}
    \label{eq: TRS}
    \mathcal{T}h^*_{\pos\pos'}\mathcal{T}^{-1} = h_{\pos\pos'}, \quad \text{with} \quad  \mathcal{T}=i\sigma_2,
\end{equation}
and is in symmetry class AII~\cite{Atland1997, kitaev2009, ludwig2015} since $\mathcal{T}^2 = -1$.
In the crystalline limit, $\theta \in \{0, \pi/2, \pi\}$, $\phi \in \{0, \pm \pi/2, \pi\}$ and $r_0=1$, it hosts a strong topological phase for $\lambda, \eta > 0$ and $2t<\epsilon<6t$.
We use $t=1$, $\epsilon=4$, $\lambda=1$, $\eta=1.8$ and take $r_0=1.3$ in all numerical calculations unless explicitly stated.

\begin{figure}[t!]
  \includegraphics[width=1\linewidth]{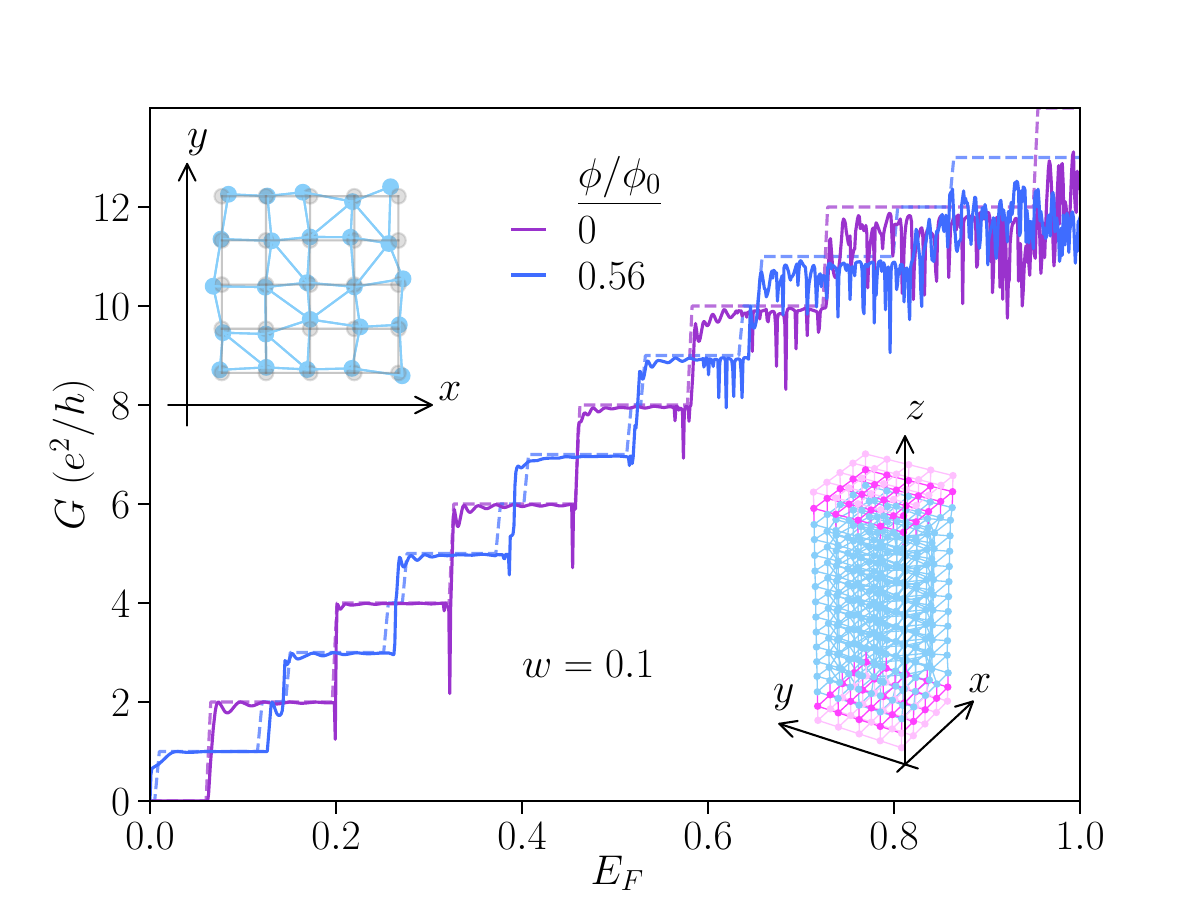}
    \caption{Conductance as a function of Fermi energy for a single layered-amorphous realization with amorphicity $w=0.1$. 
    The nanowire length is $L=200$, the dimensions of the cross section are $N_x=N_y=10$, and we set the chemical potential in the leads to $\mu=0$. 
    The conductance of the crystalline case is indicated with dashed lines. 
    The upper inset shows the amorphous cross section for a sample nanowire of dimensions $N_x=N_y=5$, with the  crystalline cross section depicted in grey. 
    The lower inset shows the full nanowire constructed from the cross section in the upper inset connected to the leads (depicted in pink).
     }
   \label{fig:AB-no-mu}
   \end{figure}

A magnetic flux threaded through the cross section of the nanowire is included in the model through Peierl's substitution.
We fix the vector potential gauge  $\boldsymbol{A}(x, y)=-B y \boldsymbol{\hat{x}}$ so that the magnetic field $\boldsymbol{B}=B\boldsymbol{\hat{z}}=\frac{\phi}{\mathcal{A}_{\rm ref}}\boldsymbol{\hat{z}}$ is parallel to the axis of the nanowire (see the inset in Fig.~\ref{fig:AB-no-mu}), with $\phi$ the flux and $\mathcal{A}_{\rm ref}$ the cross section area of the reference cubic lattice.
The hopping amplitudes and spin-orbit coupling are modified as
\begin{equation}
\begin{split}
&f_{\pos\pos'}\longrightarrow f_{\pos\pos'} \exp\left(-\dfrac{2\pi i}{\phi_0} \int_{\pos}^{\pos'} \boldsymbol{A} \cdot \boldsymbol{dl} \right)\\
 = f_{\pos\pos'}&\exp\left( i\pi \frac{\phi}{\phi_0} \frac{(y'+y)(x'-x)}{\mathcal{A}_{{\rm ref}}} \right) \quad
 \text{for} \quad f\in \lbrace t, \eta, \lambda\rbrace.
\end{split}
\end{equation}

In order to compute the electronic transport we connect the nanowire to two semi-infinite translation-invariant leads described by the crystalline limit of Eq.~\eqref{eq: Amorphous hamiltonian}.
The leads are threaded by the same magnetic flux and have the same hopping and spin-orbit parameters as the nanowire, together with an additional chemical potential $\mu$ such that $E_F$ is replaced by $E_F - \mu$.
\footnote{
There is freedom to choose the value of the chemical potential.
A nonzero chemical potential makes the density of states change abruptly from the leads to the nanowire, increasing interference and giving a highly oscillating conductance signal, see Fig.~\ref{fig:data-translation-invariant}(a).
We may then set $\mu=0$ to have a clearer and easier to interpret conductance signal, as in Fig.~\ref{fig:AB-no-mu} and Fig.~\ref{fig:G-vs-Ef-amorphous}.
In cases where the conductance is calculated at a fixed Fermi energy, the interference effects described above do not affect the results as much, and we may then use a nonzero chemical potential to allow the leads to have a higher density of states than the nanowire, acting as metallic reservoirs. 
This is the case in Fig.~\ref{fig:data-translation-invariant}(d), Fig.~\ref{fig:Conductance-flux-amorphous}(a) and Fig.~\ref{fig:boundstates}.
}.
The zero-bias conductance of the lead-nanowire-lead system is given by the Landauer formula~\cite{datta1997}
\begin{equation}
    G=\frac{e^2}{h} \mathrm{Tr}(T^\dagger T),
\end{equation}
with $T$ the transmission matrix at the Fermi energy $E_F$, which relates the amplitudes of incoming and outgoing modes at the leads.
We numerically calculate $G$ with the software package Kwant~\cite{kwant2014}.
To relate transport properties to the topology of the nanowires we use the local markers introduced in Ref.~\onlinecite{Julia2022}, which characterize the topology of states that break translation invariance and can be applied to amorphous materials~\cite{ Julia2022, Julia2024}.

\section{layered-amorphous nanowires} 
Layered-amorphous nanowires are a useful starting point for understanding how amorphicity influences electronic transport in nanowires, as they maintain translation invariance along their axis.
As a consequence, the magnetic flux is constant along the nanowire and the surface recovers an effective time-reversal symmetry at $\phi=(2n+1)\phi_0 /2$ that results in a perfectly transmitted mode.

We demonstrate the presence of a perfectly transmitted mode by calculating the conductance of a single amorphicity realization as a function of the Fermi energy, shown in Fig.~\ref{fig:AB-no-mu}. 
In the absence of flux the conductance is zero until $E_F\simeq0.06$, indicating the presence of an energy gap.
At $\phi/\phi_0=0.56$ the conductance is quantized to $G=e^2/h$ for $E_F<0.12$, reflecting the perfectly transmitted mode.
Note that the conductance is not exactly quantized down to $E_F=0$ because the leads have a finite size gap at $E_F=0$ when $\mu=0$.
Since the surface state has a finite penetration length into the bulk, the flux it encloses deviates from the exact flux threaded through the full cross section.
As a result, the perfectly transmitted mode does not occur exactly at $\phi/\phi_0=1/2$, but approaches this value in the limit of large $N_x, N_y$
\footnote{Apart from the shift induced by the finite bulk penetration depth, the position of the peaks may also be shifted to $\phi>(2n+1)\phi_0/2$ if the amorphous cross section is smaller than reference cross section area $\mathcal{A}_{\rm ref}$, and to $\phi<(2n+1)\phi_0/2$ if the amorphous cross section is bigger. 
This effect depends on the particular amorphicity realization.}.
Increasing the Fermi energy makes the flux value at which the conductance is maximal alternate between $\phi/\phi_0=0$ and $\phi/\phi_0=0.56$, as the number of modes at the Fermi energy changes between even and odd.
The main difference from the crystalline case is the sharp conductance resonances seen especially at high Fermi energy.
These come from the interface between the wire and the leads, where the lattice structure changes from being perfectly crystalline to being layered-amorphous, allowing electrons to backscatter and interfere.
When the Fermi energy reaches the bands hosting states with a nonzero bulk contribution, scattering becomes stronger leading to an increase in interference and the upward trend in conductance gets flattened.

\begin{figure}[t!]
    \includegraphics[width=1\linewidth]{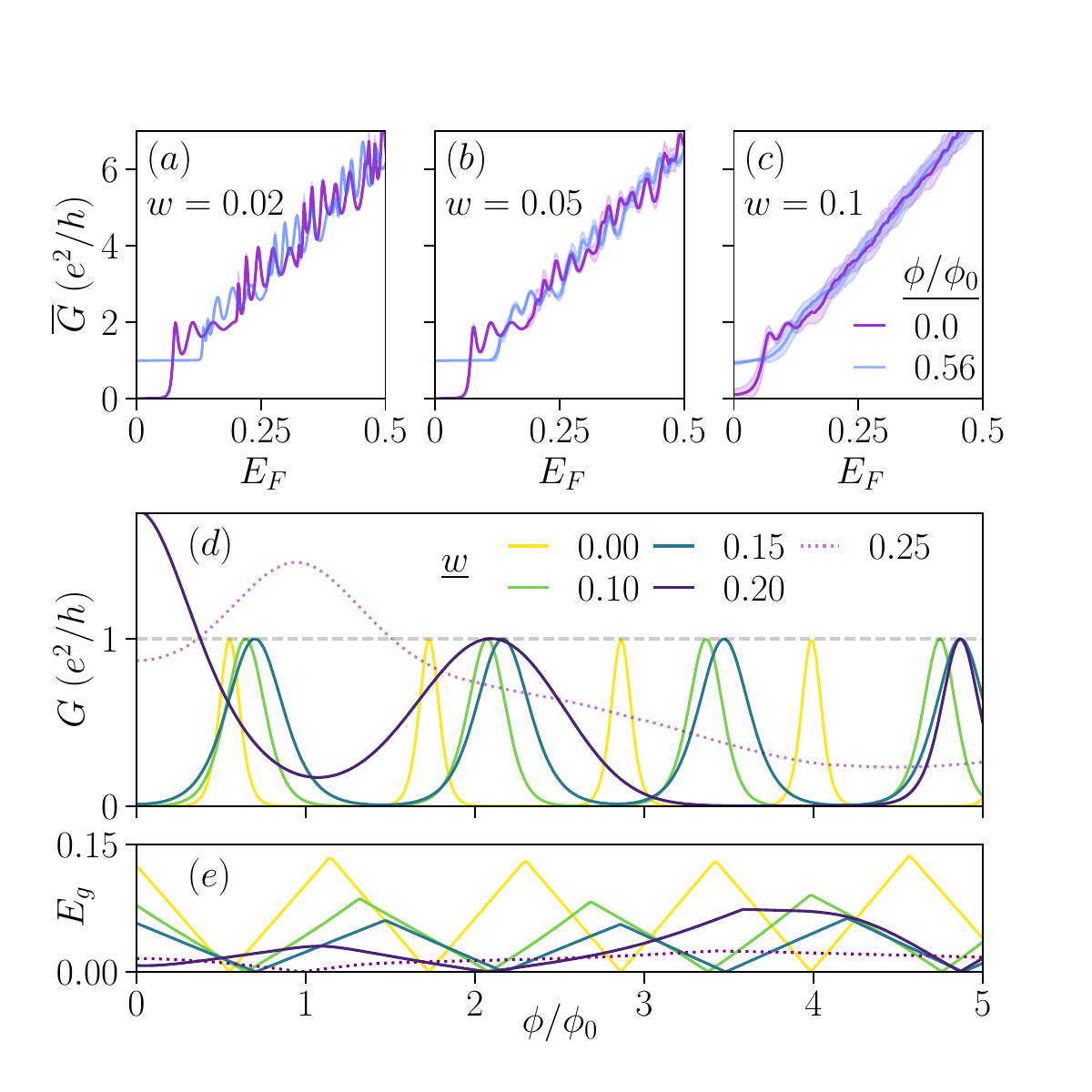}
    \caption{(a-c) Realization-averaged conductance of a layered-amorphous nanowire for three different amorphicities as a function of Fermi energy. 
    The average is performed over $300$ layer amorphicity realizations of nanowires with $L=100$ and $N_x=N_y=10$. The chemical potential in the leads is $\mu=-1$. The shaded regions indicate the standard deviation of the distribution.
    (d) Single-realization conductance at $E_F=0$ as a function of magnetic flux and for different $w$. The parameters are $\mu=-1$, $L=200$ and $N_x=N_y=10$. 
    (e) Band gap for the different nanowires in (d).}
   \label{fig:data-translation-invariant}
   \end{figure}

In Fig.~\ref{fig:data-translation-invariant}(a-c) we show the realization-averaged conductance of a layered-amorphous nanowire as a function of the Fermi energy and at three different amorphicities.
As $w$ increases, the atomic arrangements of the different nanowire realizations become more distinct from each other.
Since the exact details of each conductance realization depend on the particular nanowire structure, conductance curves dephase when averaged, leading to smoother curves than for small $w$.
The oscillations in the conductance at small $w$ are of the same nature as those at high Fermi energies in Fig.~\ref{fig:AB-no-mu}, although enhanced by the difference in chemical potential between leads and nanowire, and are present for each individual realization.
Close to $E_F=0$, the perfectly transmitted mode dominates the conductance and is not affected by interference as it cannot backscatter.
Fig.~\ref{fig:data-translation-invariant}(d) shows the single-realization conductance at $E_F=0$ as a function of magnetic flux, and Fig.~\ref{fig:data-translation-invariant}(e) the associated band gap.
Conductance peaks coincide with gap closings, deviating from the crystalline case as either $w$ or $\phi$ increase.
Despite these deviations, the conductance peaks reach perfect transmission up to $w=0.2$. 
Note that this is not the case if we calculate the  average conductance since the flux at which perfect transmission is attained changes between realizations, decreasing the height of the peak.

As amorphicity reaches $w=0.2$ the behavior of the conductance changes, exemplified by the $G>e^2/h$ peak at $\phi=0$.
This peak occurs when the gap closes while the number of modes at the Fermi energy is even.
Since these modes are not protected from backscattering, conductance is not constrained to $G=2e^2/h$.
Stronger amorphicity eventually destroys the conductance oscillations resulting in a nonzero smooth conductance as the magnetic flux is varied; see the dotted $w=0.25$ line in Fig.~\ref{fig:data-translation-invariant}(d).

\section{Fully amorphous nanowire}
\begin{figure}[t]
\includegraphics[width=1\linewidth]{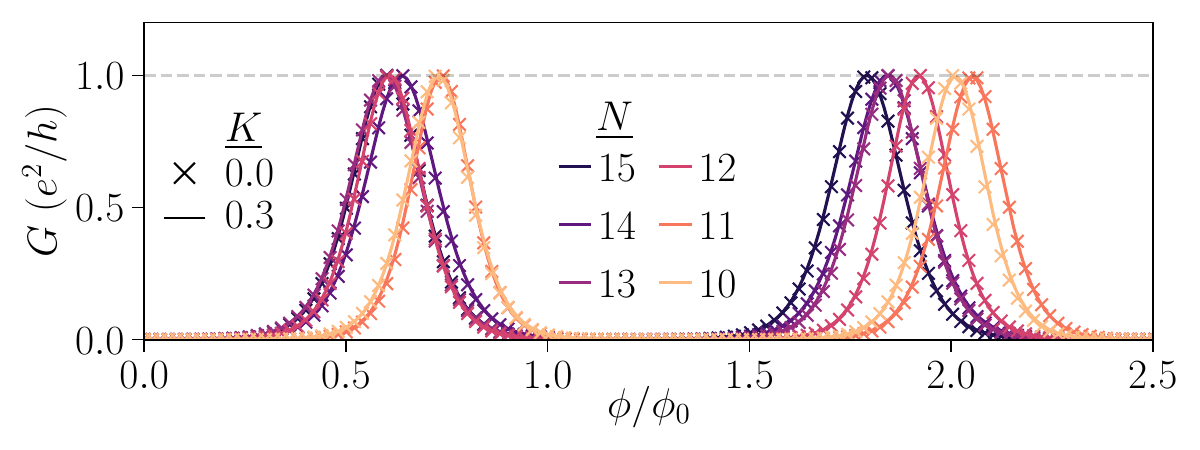}
    \caption{Effect of chiral symmetry breaking on the conductance of a single realization of a fully amorphous nanowire for different cross section linear size $N$, such that  $N_x \times N_y = N \times N$. 
    We model onsite disorder by adding the term $K_{\pos} \delta_{\pos\pos'}$, with $K_{\pos}\in[-K, K]$ drawn from a random uniform distribution for each $\pos$, to the Hamiltonian~\eqref{eq: Amorphous hamiltonian}. 
    Chiral symmetry is broken by the disorder, resulting in the conductance given by the solid lines.
    The crosses indicate the conductance for the same nanowire without onsite disorder.
    We used $w=0.15$, $L=200$ and $\mu=-1$.
    }
   \label{fig:chiral-breaking}
   \end{figure} 
Fully amorphous nanowires break translation invariance in all directions and do not have a well-defined $2d$ layer structure.
Since the area of the cross section varies along the axis, the magnetic flux is not constant along the nanowire 
\footnote{Strictly speaking the area of the cross section is not well defined, as there are no exact $2d$ layers bounding an area. }.
As a consequence, the effective time-reversal symmetry is not exact for any value of the flux and cannot protect a perfectly transmitted mode.

However, there are additional symmetries which can protect a perfectly transmitted mode, even in fully amorphous nanowires.
One possibility is chiral symmetry, which can be present at sufficiently low energies at the surface of the nanowires~\cite{Jens2020}, provided the charge-neutrality point of the surface state is not buried within the bulk bands.
Particularly in our model, chiral symmetry is not only present at the surface but it is an exact bulk symmetry, $Sh_{\pos\pos'}S=-h_{\pos\pos'}$ with $S=\sigma_3 \tau_3$, at zero Fermi energy and in the absence of scalar disorder. 
In this limit, the nanowires are in symmetry class AIII~\cite{Atland1997} and chiral symmetry protects the perfectly transmitted mode~\cite{fulga2011scattering}.

\begin{figure}[t]
\includegraphics[width=1\linewidth]{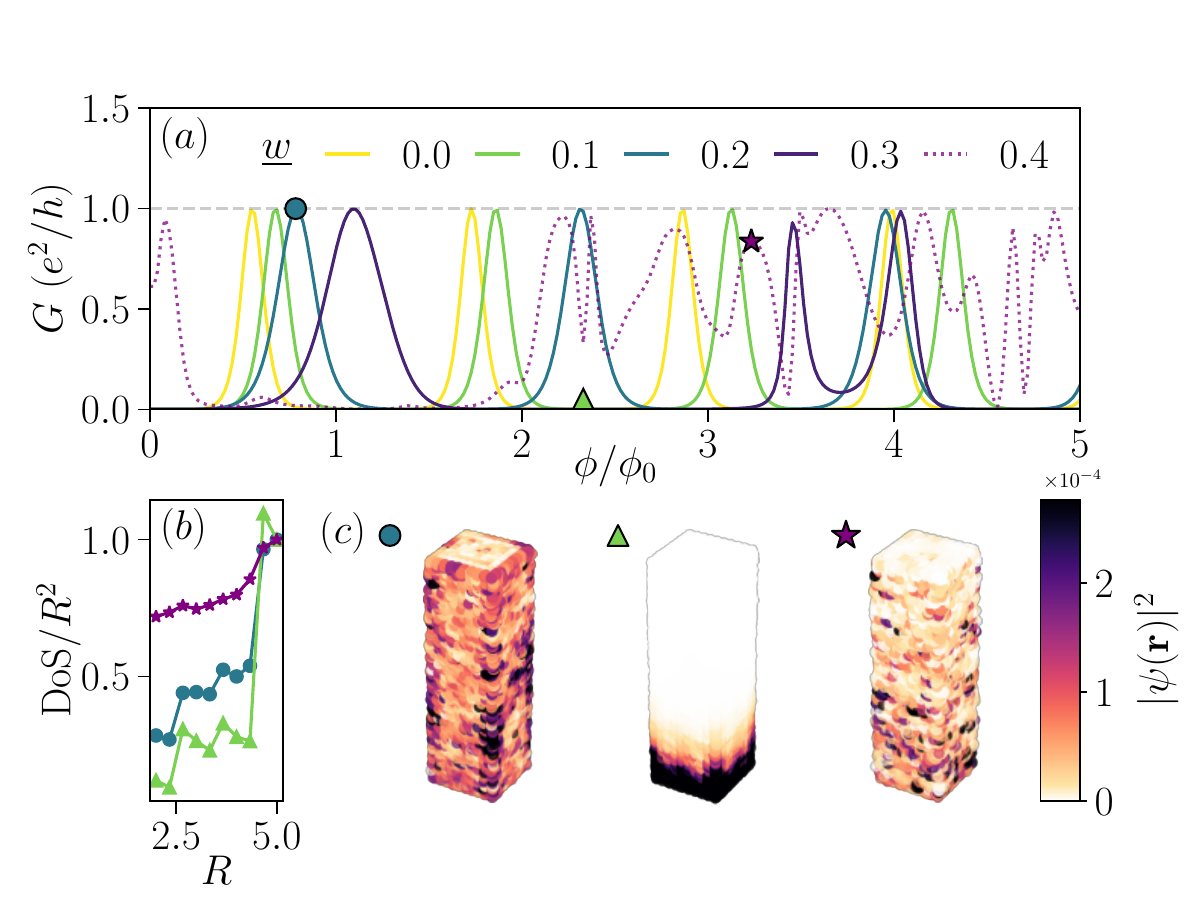}
    \caption{(a) Single-realization conductance of a fully amorphous nanowire at $E_F=0$ as a function of flux and amorphicity. The parameters are $L=200$, $N_x=N_y=10$, and $\mu = -1$. 
    (b) Total density of states per unit area as a function of the cutoff parameter $R$ for a perfectly transmitted mode (circle), localized mode (triangle) and resonant mode (star) in (a). 
    The meaning of the cutoff $R$ is sketched below the central panel in (c), where
    $R=0$ corresponds to considering only the central point in the cross section of the nanowires.
    Each curve is normalized to its total density of states per area at $R=5$.
    (c) Local density of states for the perfectly transmitted mode, localized mode and the resonant mode shown in (b).
    }
   \label{fig:Conductance-flux-amorphous}
   \end{figure} 
Even in the absence of chiral symmetry, the perfectly transmitted mode may still be protected in an average sense.
Strictly speaking, fully amorphous nanowires break time-reversal symmetry, but for a long enough nanowire the local fluctuations of the flux average out if the amorphicity is spatially isotropic.
This situation is realistic for amorphous nanowires since we expect disorder to restore inversion on average~\cite{corbae2023,Adolfo2024,spring_amorphous_2021,springMagneticAverageTI}.
Such average symmetry parallels those protecting statistical topological insulators introduced in Ref.~\onlinecite{Fulga2014}, where it was shown that the surface state remains gapless when subjected to a time-reversal-symmetry-breaking perturbation with zero average, and those protecting against localization in weak topological insulators~\cite{Jens2012,Ringel2012}.
Aside from this statistical argument, the effective time-reversal symmetry at the surface is recovered for a fully amorphous nanowire with a large enough cross section.
This follows from the fluctuations of the  flux $\delta \phi$ approaching zero for increasing linear size $N$ of the cross section, as the fluctuations of the amorphous cross section area scale with the perimeter $\delta \mathcal{A} \sim N$  while the area scales as $\mathcal{A} \sim N^2$.
Hence, $\delta A/A=\delta \phi/\phi\longrightarrow0$ as $N$ grows.

These symmetry arguments suggest that, in the regime where conductance is dominated by the perfectly transmitted mode, we expect no fundamental differences between the analysis with or without chiral symmetry 
\footnote{In order for conductance to be dominated by the perfectly transmitted mode the nanowires must be in the single mode regime which implies $E_F\sim0$, and their aspect ratio $(N_x+N_y)/L$ needs to be small enough for evanescent modes not to significantly contribute to conductance at low energies. 
Moreover, strong scalar disorder introduces energy levels within the surface gap, obscuring the Aharanov-Bohm oscillations.
}.
We  numerically support this argument by comparing the conductance in the chiral limit with the case where chiral symmetry is broken by weak onsite disorder in Fig.~\ref{fig:chiral-breaking}.
Hence, in the following, we analyze the transport in fully amorphous nanowires in the chiral limit unless otherwise stated.
\begin{figure}[]
\includegraphics[width=1\linewidth]{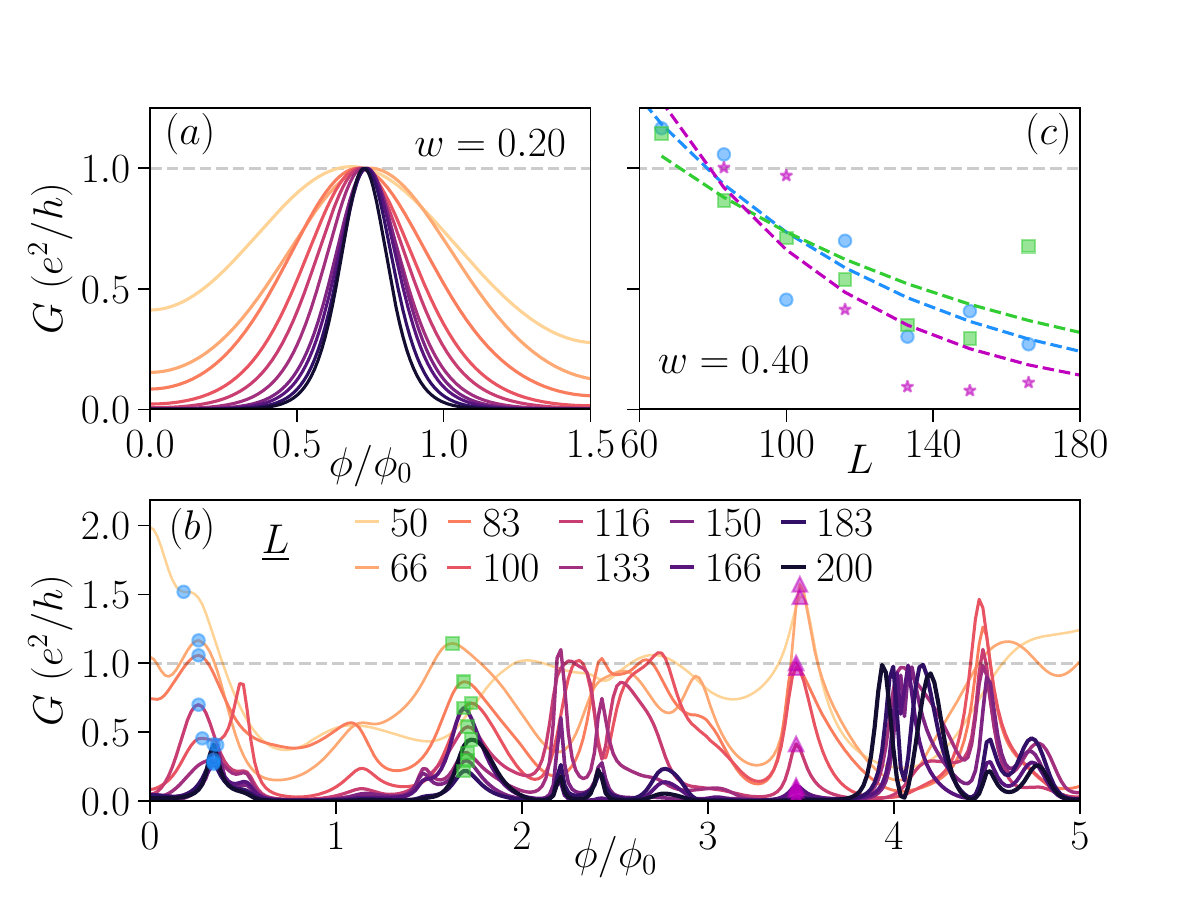}
    \caption{ Single-realization conductance of a fully amorphous nanowire at $E_F=0$ as a function of flux and nanowire length $L$, with  $w=0.2$ in (a) and $w=0.4$ in (b).
    Nanowires with different length are taken as cuts of the same parent nanowire of cross section dimensions $N_x=N_y=10$ and $\mu=-1$. 
    (c) Single-realization conductance as a function of length for the resonances indicated in (b) with a blue circle, green square and purple triangle. 
    The dashed lines show a decaying exponential fit for the three families of resonances.}  
   \label{fig:boundstates}
   \end{figure}

The single-realization conductance at $E_F=0$ as a function of flux is shown in Fig.~\ref{fig:Conductance-flux-amorphous}(a) for different amorphicities.
We can identify two different regimes.
First, for $w\lesssim 0.3$ the conductance is still dominated by the perfectly transmitted mode, protected by chiral symmetry.
The only effect of amorphicity is to broaden the conductance peaks and shift the values of flux at which they appear from the crystalline case.
At  $w\sim 0.3$ some peaks stop reaching unit conductance and for $w\sim 0.4$ the behavior has qualitatively changed to a pattern of various resonant peaks bounded, but not constrained, to $G=e^2/h$.
The change in behavior suggests that the bulk topological phase of the nanowire is lost due to the strong amorphicity.
In this regime, conductance peaks are caused by resonant transmission through bound states appearing within the surface gap.
These occur at $E\simeq 0$ and at particular values of the flux depending on the exact realization.
\begin{figure}[t!]
\includegraphics[width=1\linewidth]{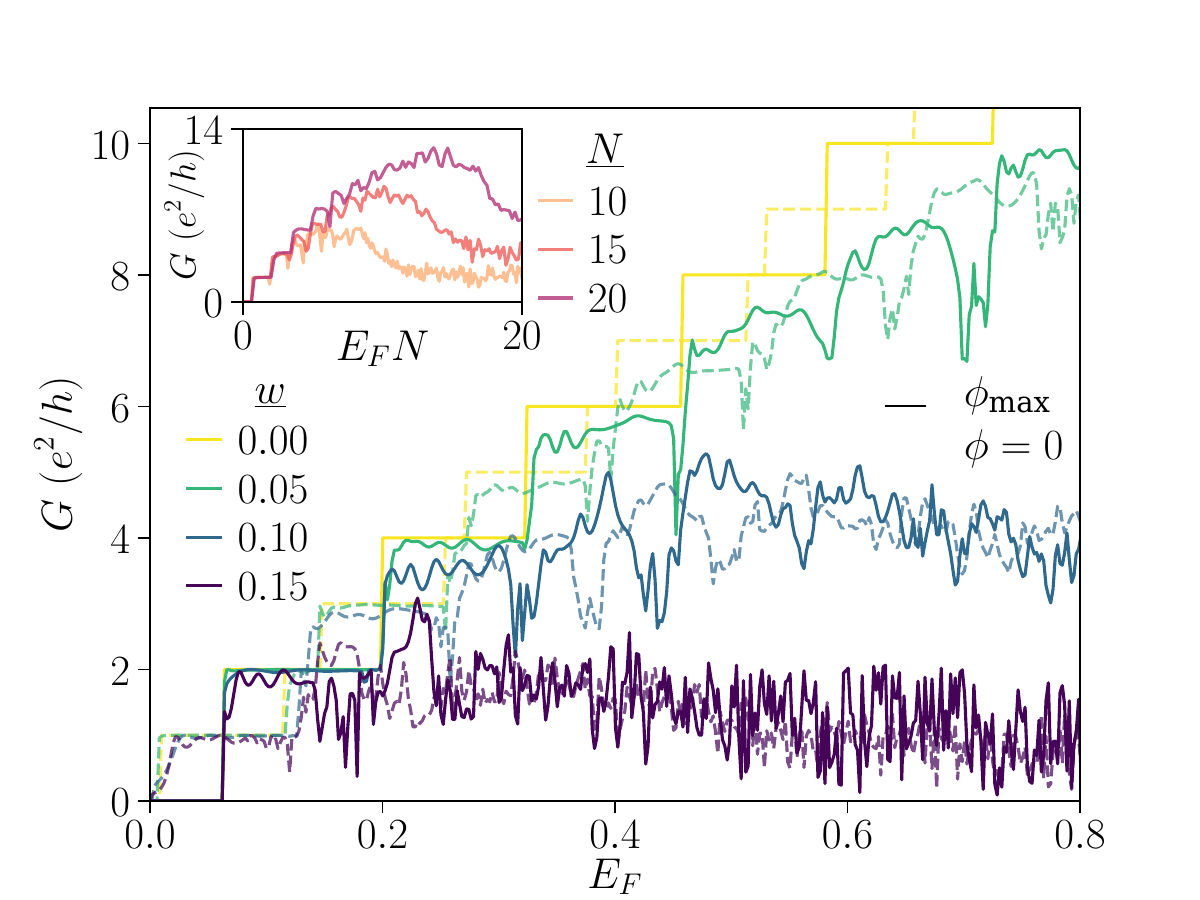}
    \caption{Single-realization conductance of a fully  amorphous nanowire as a function of Fermi energy and amorphicity. The parameters are $L=200$, $N_x=N_y=10$, $\mu=0$. The solid lines indicate the conductance at zero magnetic flux while the dotted lines indicate the conductance at the magnetic flux that maximizes conductance at $E_F=0.01$.
    The inset shows the downturn of the $\phi=0$ conductance curves for different cross section sizes as the Fermi energy reaches the bands where the localization length becomes comparable to the cross section linear size, $N$.
    The nanowires have $N_x=N_y=N$ and $L=100$.
    In the $x$-axis, $E_F$ is weighted by $N$ to account for the change in level spacing for different cross section sizes.
    }
   \label{fig:G-vs-Ef-amorphous}
   \end{figure} 

Fig.~\ref{fig:Conductance-flux-amorphous}(b) illustrates the density of states per unit area for three representative cases: a perfectly transmitted mode, a localized mode and a resonant mode. 
These examples are taken from the data in Fig.~\ref{fig:Conductance-flux-amorphous}(a), respectively marked as a circle, triangle and star.
We take regions delimited by $0<z<L$ and $\vert x - x^0\vert, \vert y-y ^0\vert<R$, with $(x^0, y^0)$ the cross section center, and increase $R$ gradually until it covers the entire cross section of the nanowire.
Note that $R=0$ corresponds to the center of the bulk, while $R=5$ corresponds to covering the entire cross section, as $N_x=N_y=10$.
The density of states in the perfectly transmitted and localized modes is mostly located in the surface, while in the resonant mode the penetration length into the bulk becomes larger due to the strong amorphicity.
The corresponding local density of states is studied in Fig.~\ref{fig:Conductance-flux-amorphous}(c).
The perfectly transmitted mode fully hybridizes with both leads without decaying, as backscattering is forbidden by the chiral symmetry.
There is an increased local density of states at the hinges of the nanowire, which is expected for topological states~\cite{Ezawa18} and it is induced because the sites in these regions have a lower coordination number than in other parts of the surface.
The localized mode  decays through the nanowire not hybridizing with the upper lead, which causes a decrease in conductance.
Finally, the resonant state hybridizes with the upper lead more faintly than the perfectly transmitted mode, resulting in  $G<e^2/h$.
The local density of states of the resonant state is inhomogeneous along the nanowire, accumulating in the darker regions, and there is no increase along the hinges, contrasting with the perfectly transmitted mode.

The difference between the perfectly transmitted mode at low amorphicity and the conductance resonances at high amorphicity is further illustrated in Fig.~\ref{fig:boundstates}.
We plot the single-realization conductance as a function of flux and nanowire length for $w=0.2$ and $w=0.4$.
We use a parent nanowire of length $L=200$ and take cuts of varying length.
This allows us to keep track of where the resonances appear, as their position is realization dependent.
Fig.~\ref{fig:boundstates}(a) shows that the conductance of the perfectly transmitted mode stays pinned to $G=e^2/h$ regardless of the nanowire length, indicating that the mode is never localized.
The broadening of the peak for shorter lengths is due to the evanescent modes of the leads that tunnel through the nanowire with a decay length that scales as $\ell_{\rm tunnel} \sim N/L$~\cite{Tworzydlo2006subpoissonian}, contributing to the conductance.
In contrast to the perfectly transmitted mode, Fig.~\ref{fig:boundstates}(b) shows that, at high amorphicity, the height of most resonances decreases with the nanowire length.
Multiple evanescent modes may contribute to the conductance in this regime, especially for short nanowires, where the conductance of certain resonances becomes $G>e^2/h$.
Fig.~\ref{fig:boundstates}(c) shows an approximate exponential decay of three different resonances in Fig.~\ref{fig:boundstates}(b), indicating the existence of bound states in the wire.

Fig.~\ref{fig:G-vs-Ef-amorphous} shows the single-realization conductance as a function of Fermi energy in the regime of low $w$, both at zero flux and at the flux value $\phi_\mathrm{max}$ that maximizes conductance at $E_F\simeq0$.
At low Fermi energy the conductance shows the same behavior regardless of amorphicity, dominated by the perfectly transmitted mode.
As in the crystalline and layered-amorphous case, the two values of the flux allow for a different number of modes at the Fermi level, resulting in the flux that gives maximal conductance alternating with the Fermi level. 
For increasing amorphicity the conductance curves start to develop a downward turn as the Fermi energy increases.
This occurs because the modes close to the band edge start populating the bulk of the nanowire, as their penetration depth into the bulk becomes comparable to the cross section linear size.
The bulk contribution induces a destructive interference and gets more pronounced with increasing amorphicity, increasing disorder and nanowire length (data not shown).
This is a crossover from surface-dominated transport to bulk-dominated transport that gives rise to Anderson localization.
The inset demonstrates that the number of modes in the window of surface-dominated transport increases with increasing cross-section linear size $N$, as the level spacing decreases with $1/N$. 
%

\section{Topology of amorphous nanowires}
Both the Aharanov-Bohm oscillations and the perfectly transmitted mode are manifestations of the topological phase of the nanowires.
The change in behavior seen in transport, losing the conductance oscillations in place of bound state resonances, thus hints at a topological transition.
In this section we explicitly characterize the topology of fully amorphous nanowires and relate it to the transport characteristics studied in previous sections.

We expect the topology of the amorphous nanowires to be protected by the approximate time-reversal symmetry in the average sense discussed in the previous section.
However, to characterize the topology of the low-energy regime where the perfectly transmitted mode stands out, we can restrict to the chiral limit.
Since the topological protection comes from time-reversal symmetry, adding a time-reversal symmetric perturbation that breaks chiral symmetry, such as adding weak disorder, will not result in a transition to a trivial phase, only in a change of the symmetry class.
Chiral symmetry acts then only as an additional symmetry that facilitates the topological characterization of the nanowires, as it makes the computation of a local topological marker more direct.
\begin{figure}[]
\includegraphics[width=1\linewidth]{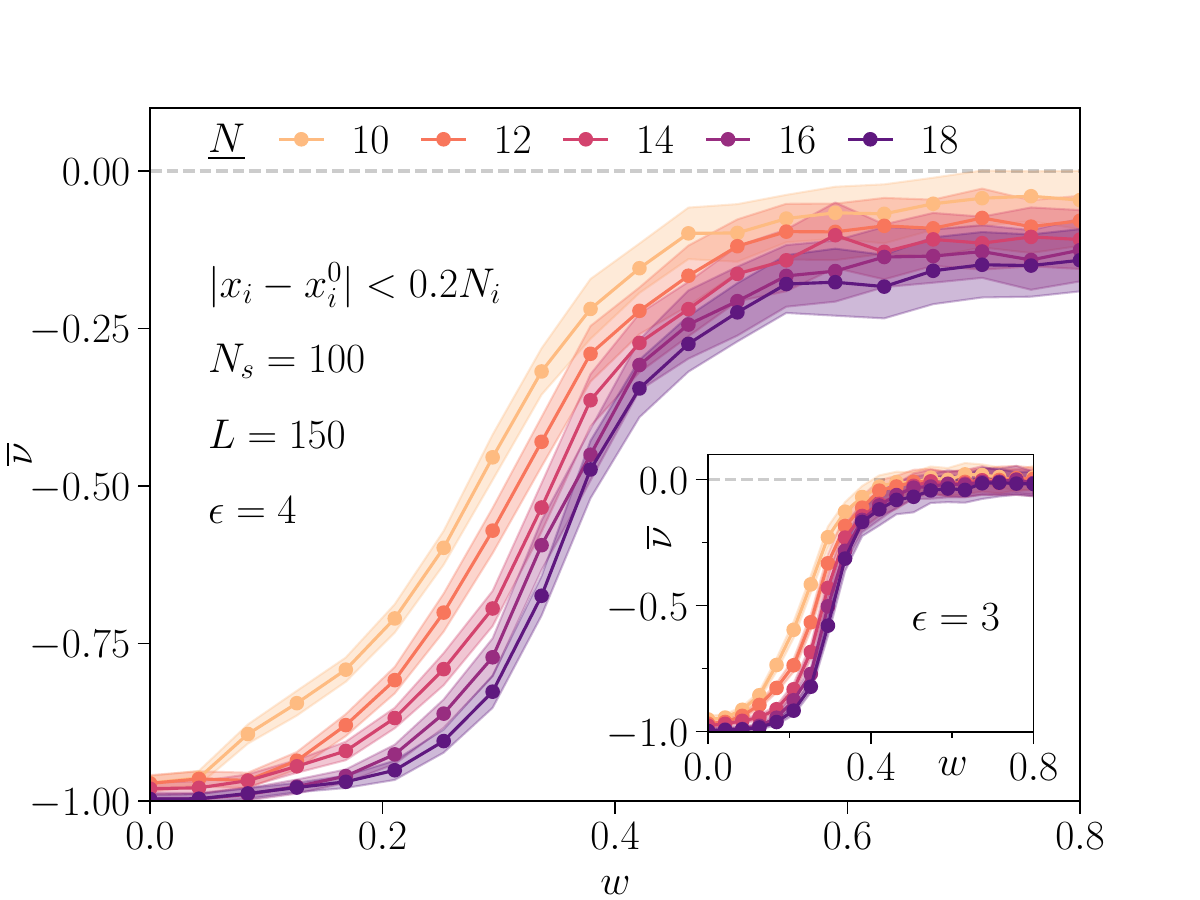}
    \caption{Realization- and bulk-averaged chiral marker for fully amorphous nanowire as a function of amorphicity and cross section size $N$, with $L=150$. 
    For each realization the local marker is averaged within a region centered at $x_i^0=(N_i -1)/2$ and spanning $\vert x_i - x_i^0\vert < 0.2 N_i$ with $x_1=x, x_2=y$ and $x_3=z$, and over $N_s=100$ realizations. 
    The shaded regions indicate the standard deviation of the distribution.
    We used $5000$ moments to calculate the one-particle density matrix with the Kernel Polynomial Method and $5$ random vectors to evaluate the average local marker with the stochastic trace algorithm.
    The inset shows an analogous calculation where we set $\epsilon=3$ (in the main panel $\epsilon=4$).}
   \label{fig:marker1}
   \end{figure} 
\begin{figure*}[t!]
\includegraphics[width=1\linewidth]{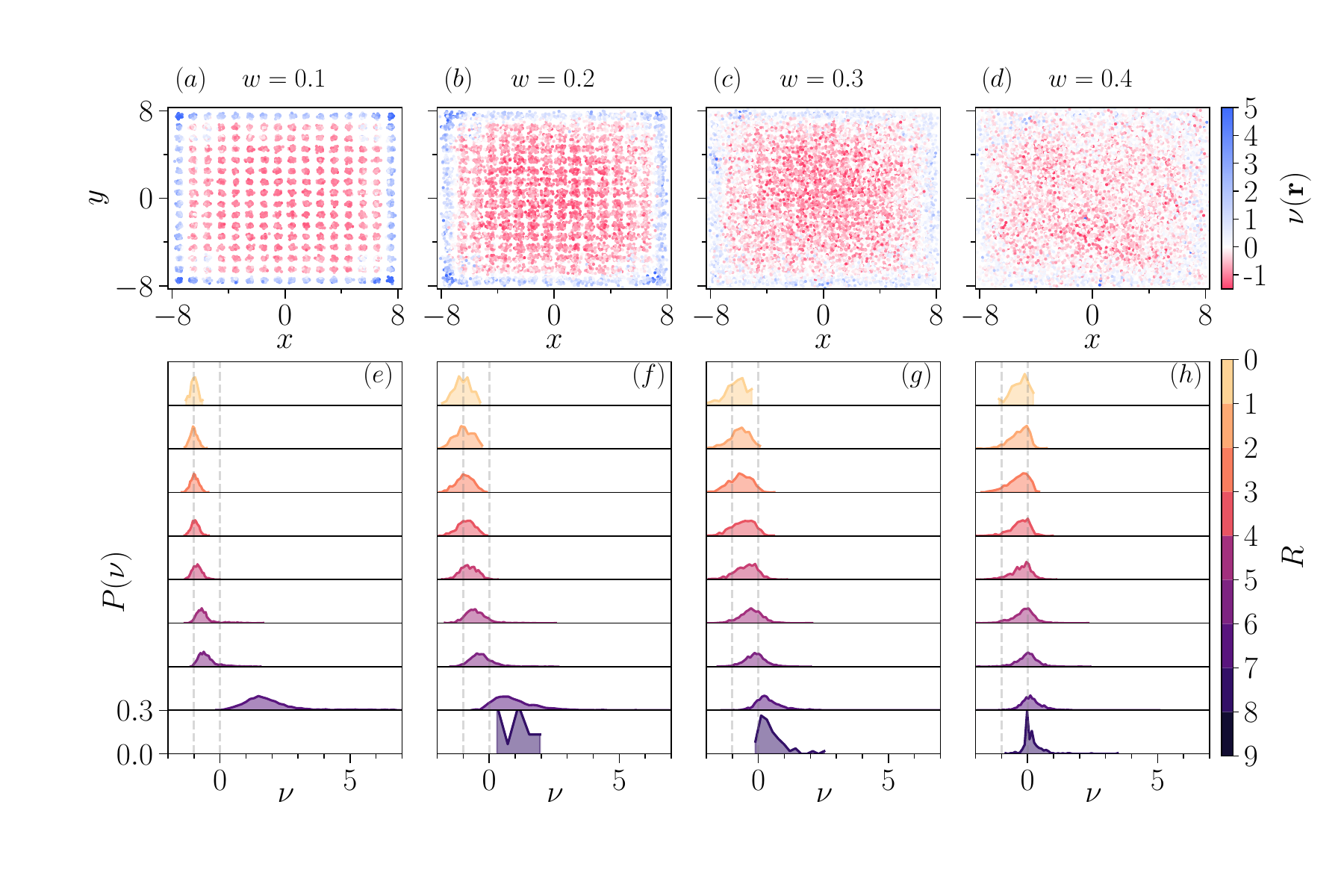}
    \caption{
    (a-d) Projection onto the $xy$-plane of all sites within a cut spanned by $40.5<z<47.5$ of a fully amorphous nanowire of length $L=100$ for different amorphicities.
    Each projection shows $7$ different realizations simultaneously, and the color map indicates the value of the chiral local marker $\nu(\pos)$ at each site.
    (e-f) Probability distributions for a site located within an inner and an outer concentric square in the $xy$-plane having a marker value of $\nu$. 
    The parameter $R$ controls the size of the inner and outer squares as in Fig.~\ref{fig:Conductance-flux-amorphous}, and the color code indicates within which concentric squares sites contribute to $P(\nu)$.
    The distributions are extracted from the same data shown in the $xy$-projections.
    The amorphicity values are (a, e) $w=0.1$, (b, f) $w=0.2$, (c, g) $w=0.3$, (d, h) $w=0.4$.
    We used $1000$ moments to evaluate the one-particle density matrix with the Kernel Polynomial method.
    }
   \label{fig:marker2}
   \end{figure*} 
In the chiral limit, fully amorphous nanowires fall in class AIII, and to characterize their topology we use the local chiral marker of Ref.~\onlinecite{Julia2022}, given by
\begin{equation}
\label{eq: local-marker}
    \nu(\pos)=-\frac{8\pi i}{3}\epsilon^{ijk}\sum_\alpha\bra{\pos, \alpha}\rho S X_i\rho X_j\rho X_k\rho\ket{\pos, \alpha},
\end{equation}
where $\epsilon^{ijk}$ is the Levi-Civita tensor, $\{\alpha\}$ denotes the set of internal quantum numbers such as spin or orbital degrees of freedom, $S$ is the chiral symmetry operator, $X_{i=1,2,3}$ is the $i$th component of the position operator and $\ket{\pos, \alpha}$ denotes a state localized at position $\pos$ and with internal quantum numbers $\alpha$.
The one-particle density matrix is  defined as $\rho_{\pos \pos'} = \bra{\psi} \hat{c}_{\pos}^\dagger \hat{c}_{\pos'}\ket{\psi}$, and carries all the information about the quantum state $\ket{\psi}$ to be characterized.
Local markers are real-space formulations of topological invariants, usually calculated by computing local expectation values of a polynomial of the one-particle density matrix, position and symmetry operators, as in Eq.~\ref{eq: local-marker}.
Averaging a local marker over a sufficiently large bulk region results in a quantized value that characterizes the topological phase of the state.
If instead we consider a full system with open boundary conditions, the average local marker over the system necessarily yields zero, as it can generally be expressed as the trace of a commutator, which vanishes in a finite Hilbert space.
In the following, we use the Kernel Polynomial Method~\cite{Weissner2006, Varjas2020} to evaluate the one-particle density matrix of the nanowires and compute the associated chiral local marker.
We combine this method with the stochastic trace algorithm~\cite{Weissner2006, Varjas2020} when performing averages over large regions of the nanowire.

Fig.~\ref{fig:marker1} shows the realization- and bulk-averaged marker $\overline{\nu}$ as a function of amorphicity and cross section size for fully amorphous nanowires of length $L=150$ and $N_s=100$ realizations. 
We also show the standard deviation of the marker distribution obtained for the different amorphous realizations.
For each realization we average the local marker over all sites within a region centered at $x_i^0=(N_i -1)/2$ and spanning $\vert x_i-x_i^0 \vert < 0.2 N_i$ with $x_1=x, x_2=y$ and $x_3=z$.
In the crystalline limit, when $w$ is close to zero, the chiral marker approaches the quantized value $\overline{\nu}=-1$ for increasing cross section size, which is expected since for the parameters chosen the crystalline nanowires are in a topological phase.
In this limit different realizations of the amorphous structure are very similar to each other, and the only source of statistical uncertainty is in the stochastic trace evaluation.
For  $w\gtrsim 0.1$ the marker starts deviating from the quantized value and the standard deviation becomes broader.
This indicates a topological transition away from the $\overline{\nu}=-1$ phase driven by amorphicity, similar to the disorder- and impurity-driven transitions studied in crystalline structures~\cite{Assuncao2024phase, Oliveira2024robustness}.
The transition results in the average marker approaching a noninteger value $\overline{\nu}\approx -0.15$ for $w\gtrsim0.6$ and increasing cross section size.
Presumably, either the single particle density matrix is not exponentially localized in this regime or the cross section size $N$ is shorter than the localization length, making the topological phase of the nanowires not well defined and resulting in a larger statistical uncertainty than at low $w$.
We stress that this situation is not an artifact of our calculation and holds physical meaning:
due to their finite cross section, nanowires may not have a well defined topological phase even if the latter is well defined in the thermodynamic limit.
These finite size effects are in-built and shape the physics of the nanowires, rather than only being of numerical importance.
We have further checked (data not shown) that the approach to a non-well-defined topological phase does not change by fixing the aspect ratio of the nanowires, ruling out the possibility of a dimensional crossover effect obscuring the results.
Although local marker calculations cannot give more information about this phase at high amorphicity, the analysis of the conductance resonances in Fig.~\ref{fig:boundstates}(b,c) suggests that the nanowires flow to a trivial insulating phase with an exponentially decaying conductance with the nanowire length, which supports the idea that the localization length of the single particle density matrix exceeds the size of the cross section.
Whether the amorphicity-driven transition converges to a well-defined trivial phase depends on the exact parameters of the calculation.
We exemplify this in the inset of Fig.~\ref{fig:marker1}, where changing the parameter $\epsilon$ results in a topological transition that converges to a trivial phase $\overline{\nu}=0$ at large $w$.
In this particular case, the bulk band gap in the crystalline limit is reduced as compared to the main panel in Fig.~\ref{fig:marker1}, which may explain why the range of amorphicities considered is enough to trivialize the nanowires as opposed to the main panel.

Complementing the average calculation in Fig.~\ref{fig:marker1}, the real space distribution of the local marker also signals a topological transition in the nanowires.
In Fig.~\ref{fig:marker2}(a-d) we show the projection onto the $xy$-plane of the sites contained in $7$ realizations of a cut spanning several cross sections (in the reference lattice) of a nanowire.
The color of each site corresponds to $\nu(\pos)$, with $\pos$ the position in the $xy$-plane.
For each projection, we compute the probability distribution $P(\nu)$ of a site situated within an inner and an outer concentric square having a certain $\nu$.
We vary the sizes of the squares, controlled by the $R$ parameter defined in Fig.~\ref{fig:Conductance-flux-amorphous}, in order to obtain different probability distributions associated with different regions in the $xy$-plane, from regions including sites very close to the center of the cross section to regions including sites only around the boundary of the nanowires.
The resulting probability distributions are shown in Fig.~\ref{fig:marker2}(e-h).
Each vertical panel in the figure corresponds to a different amorphicity.
For $w=0.1$ and $w=0.2$ the real space distribution of the marker is the one expected in a topological phase.
The nanowires have a topological core with $\nu(\pos)\approx-1$, while the boundary compensates the bulk contribution with $\nu(\pos)>0$.
This is explained more thoroughly by the probability distributions $P(\nu)$, which peak at $\nu=-1$ up to $R\simeq5$ for both $w=0.1$ and $w=0.2$.
Although peaked at $\nu=-1$, the probability distributions for small $R$ broaden with increasing amorphicity.
At $R\gtrsim5$ for $w=0.1$ and $w=0.2$ the distributions start to drift towards higher marker values, eventually peaking and spreading over positive values of $\nu$.
The drift shows how sites at the boundary compensate for the inner topological core, making clear the distinction between bulk sites and boundary sites.
For $w=0.3$ the peaks in $P(\nu)$ are not pinned to $\nu=-1$, and although there is a drift, the distributions for the boundary sites generally peak at $\nu=0$ and spread to both negative and positive values.
This signals that $\nu(\pos)$ in the bulk is not as homogeneously quantized as for smaller amorphicity, and hence induces less compensation by the boundary.
At $w=0.4$ all $P(\nu)$ peak around $\nu=0$.
The distributions spread mainly to negative values for small $R$, consistent with $\overline{\nu}$ staying below zero for the range of amorphicities considered in Fig.~\ref{fig:marker1}, while
they spread both to positive and negative values for larger $R$.
Although the positive values are generally localized close to the surface, the distinction between the bulk and the boundary becomes faint, as negative and positive values of $\nu(\pos)$ become less probable than $\nu=0$, see Fig.~\ref{fig:marker2}(d), indicating a topological transition.

From this analysis, we conclude that the sharp oscillations in conductance appear when the topological protection of the perfectly transmitted mode is lost due to the transition away from the topological phase driven by amorphicity.

\section{Discussion}
We have analyzed the transport properties of both layered-amorphous and fully amorphous topological insulator nanowires.
The low-energy properties up to moderate amorphicity are dominated by the presence of a perfectly transmitted mode in both models, as in crystalline nanowires.
In the layered-amorphous case, the perfectly transmitted mode is protected by the effective time-reversal symmetry of the surface and is progressively lost for increasing amorphicity.
The resulting regime shows a smooth behavior of the conductance as the magnetic flux varies.
In the fully amorphous case, the perfectly transmitted mode can be protected by chiral symmetry when $E_F=0$ and in the absence of scalar disorder, as well as by statistical time-reversal symmetry for long and wide enough wires.
For strong amorphicity the conductance shows a proliferation of sharp resonances, in contrast to the smooth behavior in the layered-amorphous case.
These resonances follow an approximate exponential decay with the nanowire length and are attributed to the formation of bound states.

Using local topological markers, we have seen that the conductance resonances act as transport signatures of a phase transition away from a topological phase.
In particular, amorphicity drives the nanowires from a topological phase characterized by a quantized chiral marker $\nu=-1$ to an insulating phase where the topology may not be well defined.
Although our calculations in the fully amorphous case focus on the chiral limit, we argued that our results, both in transport and topological characterization, are representative for the regime where the perfectly transmitted mode is the most relevant contribution to conductance, even in the absence of chiral symmetry.

With the parameters chosen in our calculations, the clean bulk band gap is of the order of $\Delta E \sim 2t$, with a Fermi velocity of $v_F \sim 1.7$, giving a localization length for the surface states in a clean nanowires of $\xi=v_F/ \Delta E\sim 1$ in units of the lattice spacing.
In our calculations $N$ is sufficiently large compared to $\xi$ for finite size effects to be suppressed in the clean and low amorphicity regime.
For the topological transition to occur, the mobility gap should close increasing the localization length to values larger than the cross section. Hence, at large amorphicities the marker may not accurately capture the topological phase depending on the chosen parameters, explaining the lack of quantization of the marker at high amorphicity.

The nanowires we have used are also accessible in experiments, since amorphous nanowires with a smooth and approximately constant cross section may be achieved by FIB etching a nanowire~\cite{hunter2024, estry2024}.
Experimentally, the magnetic field needed to see the phenomena we describe is the same for crystalline and amorphous nanowires, and depends on the cross section size as the flux needs to be of the order of one flux quanta.
The dimensions of experimental nanowires vary depending on the particular work, but their perimeter ranges usually from $100-450$nm such that the magnetic field required to get $\phi=h/e$ ranges from $0.3-10$T~\cite{hong2014}.
Although in our calculations we cannot reach experimentally relevant sizes, the behavior we see is relevant for experiments as long as the level broadening is smaller than the level spacing, as is our case.

Our results contribute to understanding how transport and local topological markers can be used as a probe for amorphous topological states, and explain how certain amorphous structures and symmetries, including statistical symmetries~\cite{Fulga2014,corbae2023,spring_amorphous_2021,springMagneticAverageTI}, may be important in order to create amorphous topological nanowires for further applications.
The methods we used could also be potentially useful to explain the transport properties of other amorphous materials, such as for example the anomalous resistivity scaling with thickness seen in thin films of amorphous CoSi~\cite{rocchino2024}.

\section{Acknowledgments}
We thank T.~Klein~Kvorning, J.~D. Hannukainen, L.~Gómez-Paz and I.~M.~Flór for helpful and insightful discussions.
This work received funding from the European Research Council (ERC) under the European Union’s Horizon 2020 research and innovation program (Grant Agreement No. 101001902), The Knut and Alice Wallenberg Foundation (KAW) via the project Dynamic Quantum Matter (2019.0068), and from the Swedish Research Council (VR) through Grant No. 2020-00214.
A. G. G. is supported by the European Research Council (ERC) Consolidator grant under grant agreement No. 101042707 (TOPOMORPH).
The computations were partially enabled by resources provided by the National Academic Infrastructure for Supercomputing in Sweden (NAISS) at the National Supercomputer Centre at Linköping University partially funded by the Swedish Research Council through grant agreement no. 2022-06725. 

\nocite{zenodo-repo}
\bibliography{refs}
\end{document}